\documentclass[preprint]{revtex4-2}
\usepackage{
  color,
  graphicx,
  amsfonts,
  amsmath,
  float,
  physics,
  siunitx,
  ulem
}
\usepackage{xspace}
\usepackage[version=4]{mhchem}
\newcommand{\fref}[1]{Fig.~\ref{#1}}  
\renewcommand{\eqref}[1]{eq.~\ref{#1}} 

\newcommand{\CEP}{\ensuremath{\phi_{\mbox{\small \sc ce}}}\xspace}

\newcommand{\parallelsum}{\mathbin{\!/\mkern-5mu/\!}}

\usepackage[utf8]{inputenc}
\usepackage[T1]{fontenc}
\usepackage[final]{changes}

\newcommand{\LCPMR}{Sorbonne Universit\'e, CNRS, Laboratoire de Chimie Physique--Mati\`ere et Rayonnement, 75005 Paris, France}

\begin{document}
\title{Vibronic correlations in molecular strong field dynamics.}

\author{Marie Labeye}
\affiliation{PASTEUR, D\'epartement de Chimie, \'Ecole Normale Supérieure, PSL University, Sorbonne Université, CNRS, 75005 Paris, France}
\email{marie.labeye@ens.psl.eu}
\author{Camille Lévêque}
\author{François Risoud}
\author{Alfred Maquet}
\author{J\'er\'emie Caillat}
\author{Richard Ta{\"\i}eb}
\affiliation{\LCPMR}
\begin{abstract}
We investigate ultrafast vibronic dynamics triggered by intense femtosecond infrared pulses in small molecules. Our study is based on numerical simulations performed with 2D model molecules, and analyzed in the perspective of the renown Lochfrass and Bond-Softening models. We give a new interpretation of the observed nuclear wave packet dynamics, with a focus on the phase of the bond oscillations. Our simulations also reveal intricate features in the field-induced nuclear motion that are not accounted for by existing models. 
Our analyses assign these features to strong dynamical correlations between the active electron and the nuclei, which significantly depend on the carrier envelope phase of the pulse, even for relatively ``long'' pulses, which should make them experimentally observable.
\end{abstract}

\maketitle
\section{Introduction}
With the development of intense laser sources, it is now possible to probe ultrafast molecular dynamics at the femtosecond time scale \cite{legare_imaging_2005,alnaser_simultaneous_2005,leone_DBr} and below \cite{baker_probing_2006,baker_dynamic_2008,alnaser_subfemtosecond_2014}. While these studies were for a long time restricted to the vibrational dynamics in electronic excited states or in ionized species, femtosecond nuclear dynamics have been recorded in real time in the electronic ground state  (EGS) of several diatomic molecules, e.g. \ce{D2} \cite{ergler_quantum-phase_2006}, \ce{I2} \cite{fang_strong-field_2008,fang_comparison_2008}, and \ce{Br2} \cite{hosler_characterization_2013}, and a few years ago in larger molecules, e.g. \ce{CO2} \cite{rudenko_strong-field-induced_2016}, \ce{NH3} \cite{forster_imaging_2016}, \ce{CH3I} \cite{wei_elucidating_2017}, \ce{CH2I2} \cite{wei_ultrafast_2019}, \ce{CH3OH} \cite{ando_strong-field_2019}, \ce{H2O} \cite{kageyama_vibrational_2022}. These experiments used a femtosecond infrared pulse to create a coherent vibrational wave packet in the EGS. However such coherent population transfer raises the question of the vibrational excitation mechanism. Indeed, in the particular case of homonuclear diatomic molecules, the electric field does not couple the vibrational states within a given electronic state. Two different mechanisms were proposed, dubbed Bond-Softening (BS) and Lochfrass (LF) \cite{goll_formation_2006}. Both are based on an adiabatic approach where the electronic and nuclear dynamics are treated separately. The former, BS, is related to the instantaneous Stark shift while the latter, LF, is related to the instantaneous strong-field ionization rate. From the nuclei point of view these couplings can be seen as an effective complex potential, which depends on the internuclear distance $R$, and effectively distorts the potential energy surface (PES) of the EGS in the complex plane. The Stark shift constitutes the modification of the real part of this effective potential while the ionization rate amounts to its imaginary part, i.e., the lifetime of the dressed state~\cite{saenz_enhanced_2000,saenz_behavior_2002}. 

Although the BS and LF mechanisms occur simultaneously and may in principle interfere, they are usually thought of as being independent from each other, and are thus treated separately~\cite{ergler_quantum-phase_2006,goll_formation_2006,fang_comparison_2008}. For BS one needs to solve the Born-Oppenheimer (BO) nuclear Time-Dependent Schr\"odinger Equation (TDSE) taking into account only the Stark shift, while for LF one solves the TDSE with only the ionization rate. In the two cases one finds that, after the interaction with the pulse, some population has been transferred from the ground to the first vibrational excited state, and that the population in the higher excited states remains negligible, consistently with experimental results\cite{ergler_quantum-phase_2006}. The nuclear wave packet thus starts to oscillate at a frequency $\omega_\mathrm{vib}$ equal to the energy difference between these two vibrational states. The phase $\Phi$ of these oscillations depends on the considered mechanism and is therefore usually used to distinguish them \cite{ergler_quantum-phase_2006,fang_strong-field_2008,kageyama_vibrational_2022}. If one takes only LF into account then $\Phi_\mathrm{LF}=\pi$, while if only BS is considered then $\Phi_\mathrm{BS}=\pi/2$~\cite{ergler_quantum-phase_2006}. This phase was experimentally measured in Ref.~\citenum{ergler_quantum-phase_2006} for \ce{D2} where they obtained $\Phi_\mathrm{exp}=0.946\pi$, and in Ref.~\citenum{fang_strong-field_2008} $\Phi_\mathrm{exp}=0.81\pi$, for \ce{I2}. In both cases they concluded that it was a direct experimental proof of the observation of Lochfrass. The phase was also measured for different vibrational transitions of \ce{H2O} in Ref.~\citenum{kageyama_vibrational_2022} where the value of the phase was used to distinguish which mechanism was predominant for each transition.

However there is no established theoretical background relating the value of this phase $\Phi$ to the relative importance of LF and BS which would support such a conclusion. Besides, simulations that include only one or the other of the two mechanisms are somewhat artificial. To justify such a separation, one would have to consider that BS and LF are completely decoupled. This is far from being intuitive, since they can occur simultaneously, and both affect the nuclear dynamics. 

Moreover, the LF and BS models allow to include some of the electron-nuclei coupling but only in an averaged and adiabatic way, where the electrons adapt instantaneously both to the nuclei, and to the laser electric field. Although they successfully reproduce the phase of the nuclei oscillations in \ce{D2} \cite{ergler_quantum-phase_2006}, their full potential and limitations have never been thoroughly investigated. In particular, since they consider electrons and nuclei separately, they cannot reproduce the \textit{full} vibronic correlations that drive the observed dynamics. 

In this article we analyze the different vibronic correlations that are included in the LF and BS models and their effect on the field-induced nuclear dynamics of a low dimensional model molecule reminiscent of \ce{H2}. In section~\ref{sec:num_methods} we present our numerical framework. Section~\ref{sec:results} is organized in three parts, first we assess the qualitative relevance of approaches\cite{goll_formation_2006,fang_comparison_2008} where the LF and BS mechanisms were treated separately. We compare the predictions of these models to a complete BO model, where both mechanisms are included simultaneously\cite{ergler_quantum-phase_2006}, to question the limits of the interpretations based on the LF/BS dichotomy. Hereafter, we refer to this combined approach, first implemented in Refs.~\citenum{saenz_influence_2000} and~\citenum{ergler_quantum-phase_2006}, as LBS for Lochfrass-Bond Softening. We then compare this LBS model to fully correlated simulations where both electrons and nuclei are treated at the same level of theory. We show evidence that, later on, parts of the vibronic wave-packet leaving the EGS eventually rescatter towards the $x=0$ origin where the EGS is located, i.e., the electron returns to the nuclei. This rescattering electron dynamics manifests as interferences which consequently affect the nuclear motion in the EGS, through vibronic correlation. These highly correlated dynamics are completely absent from the LBS model and have not been characterized before. Interestingly, our numerical simulations show that it depends strongly on the carrier envelop phase (CEP) of the incident laser pulse, and that, for short pulses, this dependence counter-intuitively increases with the pulse duration. Finally we show that these findings can be reproduced with simulations performed in a BO-states basis, provided than the active space is large enough.

\section{Numerical methods}
\label{sec:num_methods}
We investigate the nuclear dynamics of a homonuclear diatomic model molecule submitted to a strong femtosecond infrared pulse. We use 2D model systems, where the electron is confined in the direction of the linearly polarized electric field, for which extensive simulations can easily be performed. We compare two different systems: one where the field is aligned with the molecular axis and one where the field is perpendicular. In this section we detail the different models used to simulate the dynamics of these model molecules. Atomic units are used throughout this article, unless otherwise stated. Since we are mainly interested in excitations from the ground to the first vibrational excited state of the EGS, all times will be expressed in units of this transition period $T_\mathrm{vib}=2\pi/\Delta E_\mathrm{vib}=\SI{8.05}{fs}$, where $\Delta E_\mathrm{vib}=\SI{0.514}{eV}=\SI{4140}{cm^{-1}}$. All simulations details are given at the end of the section.

\subsection{Laser pulse}
In all our numerical calculations, we take a sine-square envelope for the electric field:
\begin{align}\label{eqn:field}
F(t)=F_0\sin^2\!\left(\frac{\omega_\mathrm{L}}{2N_\mathrm{c}}t\right)\sin(\omega_\mathrm{L}t+\CEP),
\end{align}
where $F_0\approx\SI{0.09}{a{.}u{.}}$ is the field amplitude corresponding to a peak intensity of $I_\mathrm{L}=\SI{3e14}{W/cm^2}$, $\omega_\mathrm{L}=\SI{1.55}{eV}=\SI{12500}{cm^{-1}}$ is the field angular frequency corresponding to a \SI{800}{nm} Ti-Sa laser, $N_\mathrm{c}$ is the total number of optical cycles in the pulse set to $N_\mathrm{c}=8$ unless otherwise stated, and \CEP is
the CEP. With these parameters, the peak ponderomotive potential of the laser pulse is $U_\mathrm{p}=\SI{17.9}{eV}$.

\subsection{Two dimensional model systems}
\label{sec:2D_systems}
\begin{figure}[htbp]
    \centering
    \includegraphics[width=0.7\linewidth]{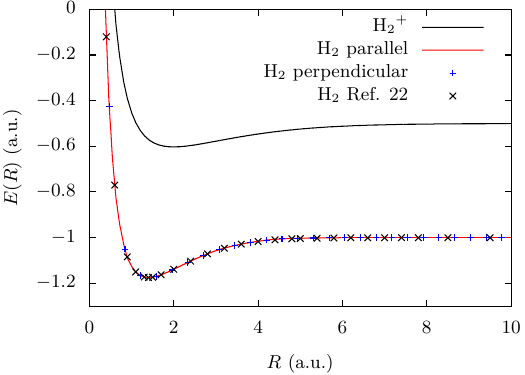}
    \caption{Field-free BO potential energy surfaces of our 2D model systems. Ground electronic state of \ce{H2+} (black) corresponding to $V_\mathrm{NN}$, electronic ground state energy of \ce{H2} of the parallel (red) and perpendicular (blue crosses) systems, and BO electronic ground state of \ce{H2} from Ref.~\citenum{kolos_new_1986} (black x).}
    \label{fig:energies}
\end{figure}

In our 2D systems, the first dimension corresponds to the electron position~$x$, and the second dimension to the internuclear distance~$R$. The total Hamiltonian reads:
\begin{align}
\label{eq:Lochfrass XR hamiltonian}
H(x,R,t)=&-\frac{1}{2\mu}{\pdv[2]{}{R}}-\frac{1}{2}{\pdv[2]{}{x}} + V_{\mathrm{NN}}(R)+ V_{\mathrm{eN}}(x,R) + H_{\mathrm{int}}(x,t)
\end{align}
where the first two terms are kinetic energy terms with $\mu$ the reduced mass of the nuclei, $V_{\mathrm{NN}}$ is the nucleus-nucleus interaction which will be taken to be equal to the PES of the electronic ground state of \ce{H2+} which was obtained by solving the time independent Schrödinger equation for the electron, through a shooting algorithm implemented in elliptical coordinates, at each internuclear distance. These numerical energies, which were previously used in a different context \cite{caillat_classical_2004}, are shown as a solid black line on \fref{fig:energies}. Finally, $V_{\mathrm{eN}}$ is the nuclei-electron interaction potential and $H_\mathrm{int}$ is the interaction with the field in length gauge:
\begin{equation}
\label{eq:lochfrass length gauge}
H_\mathrm{int}=xF(t).
\end{equation}
The nuclei-electron interaction depends on the system. For the parallel case, where the electron is confined in the direction of the molecular axis, we use a molecular Soft-Coulomb potential:
\begin{align}
\label{eq:Lochfrass molcoul}
V^{\parallelsum}_{\mathrm{eN}}(x,R)=&-\frac{0.5}{\sqrt{a(R)^2+\left(x+R/2\right)^2}} -\frac{0.5}{\sqrt{a(R)^2+\left(x-R/2\right)^2}},
\end{align}
where the regularization parameter $a(R)$ is adapted so that the electronic ground state at each value of $R$ has the same energy as the one of the real \ce{H2} molecule, within the BO approximation, as reported in Ref.~\citenum{kolos_new_1986}. The obtained values (see section~\ref{app:numerical_details}) are shown as a solid red line on \fref{fig:energies}, along with the data from Ref.~\citenum{kolos_new_1986} (black x).
For the perpendicular case, where the electron is confined in the direction perpendicular to the molecular axis, we use a simple Soft-Coulomb potential:
\begin{equation}
\label{eq:Lochfrass Coulomb}
V^{\perp}_{\mathrm{eN}}(x,R)=-\frac{1}{\sqrt{a(R)^2+x^2}},
\end{equation}
where again we fit the regularization parameter $a(R)$ so that the electronic ground state energy matches the one of the real \ce{H2} molecule, within the BO approximation, as reported in Ref.~\citenum{kolos_new_1986}. The obtained values (see section~\ref{app:numerical_details}) are shown as blue crosses on \fref{fig:energies}, along with the data from Ref.~\citenum{kolos_new_1986} (black x).

Although these are simple 2D model systems, they are, to some extent, comparable to a real \ce{H2} molecule. Indeed, the electronic dynamics will mostly happen along the polarization direction of the field. Moreover, by construction of our model, the obtained BO vibrational states and energies in the EGS of the neutral and of the ion match those of the actual H$_2$ molecule treated in the BO approximation (and ignoring rovibrational couplings). Even though, the ``parallel'' and ``perpendicular'' models have the same BO states, they are not equivalent. In particular, the parallel case has a larger polarizability, which affects both the DC Stark shift and the strong-field ionization rate \cite{labeye_tunnel_2018}, and thus, as we will see in section~\ref{sec:results}, the BS and LF mechanisms. Note that our model molecules also have electronic excited states in the neutral, that are not intended to match the exact ones. Moreover, since it is a single active electron system, there is no electronic excited states in the ion in our model: all continuum electronic states correspond to the electronic ground state of \ce{H2+}. We also expect the correlations to be overestimated by the low dimensionality \cite{henkel_interference_2011,vanne_alignment-dependent_2010}.

\subsection{Lochfrass and Bond Softening\added{ models}}
\label{sec:LBS_def}
The Lochfrass and Bond-Softening models rely on the Born-Oppenheimer approximation: the wave function is factorized in an electronic and a nuclear contribution: 
\begin{equation}
\label{eq: BO time-dependent wave function}
\Psi(x,R,t)=\widetilde{\varphi}_0(x;R,F(t))\chi(R,t),
\end{equation}
where $\widetilde{\varphi}_0(x;R,F(t))$ is the electronic ground state \textit{dressed} by the instantaneous electric field~$F(t)$, for which $R$ and $t$ are just parameters, and $\chi(R,t)$ is the nuclear wave function that propagates on this dressed electronic state. The adiabatic approximation supposes that this field-dressed state $\widetilde{\varphi}_0(x;R,F(t))$ instantaneously adapts to the value of the time-dependent electric field. It is thus solution, for each time~$t$, of the field-dressed electronic time-"independent" -- in the sense that the time is just a constant parameter -- Schrödinger equation:
\begin{equation}
\label{eq: time-dependent dressed tise}
H_\mathrm{el}(x,R,F(t))\widetilde{\varphi}_0(x;R,F(t))=\varepsilon_0\big(R,F(t)\big)\widetilde{\varphi}_0(x;R,F(t)),
\end{equation}
where
\begin{equation}
\label{eq: time-dependent dressed hamiltonian}
H_\mathrm{el}(x,R,F)=-\frac{1}{2}{\pdv[2]{}{x}}+ V_{\mathrm{Ne}}(x,R)+ xF
\end{equation}
is the field-dressed electronic Hamiltonian. The field-dressed energy can be decomposed into three terms:
\begin{equation}
\label{eq: lochfrass complex energy}
\varepsilon_0(R,F)=E_0(R) + \Delta E(R,F)-i\frac{\Gamma(R,F)}{2}.
\end{equation}
that all have meaningful physical interpretation.

The first term $E_0(R)$ is the field-free electronic energy of the neutral electronic ground state. This term directly gives the field free PES of the EGS as $V_\mathrm{PES}(R,F=0)=V_\mathrm{NN}(R)+E_0(R)$. 

\begin{figure}[htbp]
    \centering
    \includegraphics[width=0.49\linewidth]{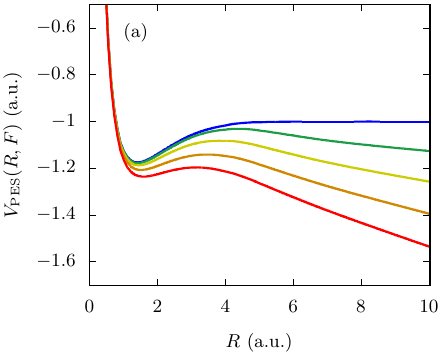}
    \includegraphics[width=0.49\linewidth]{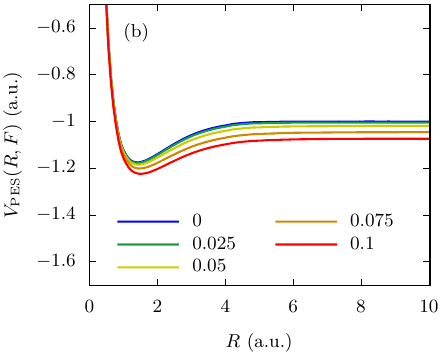}
    \caption{Stark shift of the neutral electronic ground state for our 2D systems: (a) parallel and (b) perpendicular case. The PES (\eqref{eq:PES_def}) is shown as a function of $R$ for various values of $F$ given in a.u. in the legend.}
    \label{fig:stark_shift}
\end{figure}

The second term $\Delta E(R,F)$ is the DC Stark shift, and accounts for the dressing of the state by the field \cite{bucksbaum_BS_H2+_1990}, mainly through two-photon interaction with the electronic excited states. This term will distort the real part of the effective potential as the electric field oscillates. We show in \fref{fig:stark_shift} the dressed PES
\begin{align}
\label{eq:PES_def}
    V_\mathrm{PES}(R,F)=V_\mathrm{NN}(R)+E_0(R) + \Delta E(R,F)
\end{align}
as a function of $R$ for a few values of the field $F$ computed with the $R$-box method \cite{maier_spherical-box_1980} (see section~\ref{app:numerical_details}). Consistently with previous results \cite{saenz_enhanced_2000}, we observe that, in the parallel case [\fref{fig:stark_shift}(a)], the field distorts the PES by reducing the energy barrier to dissociation, and thus soften the molecular bond. However, in the perpendicular case [\fref{fig:stark_shift}(b)] this bond-softening effect is suppressed. 

\begin{figure}[htbp]
    \centering
    \includegraphics[width=0.49\linewidth]{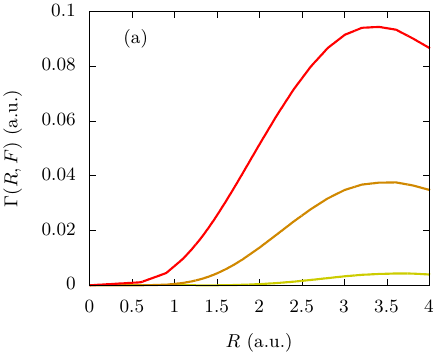}
    \includegraphics[width=0.49\linewidth]{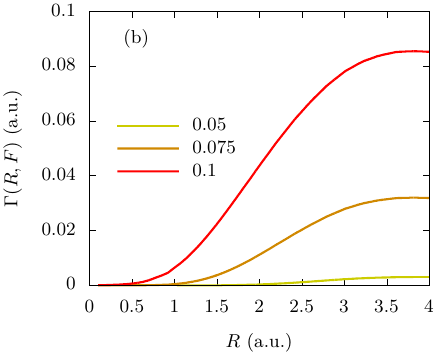}
    \caption{Strong-field ionization rate $\Gamma(R,F)$ of our 2D systems: (a) parallel and (b) perpendicular case. The rate is shown as a function of $R$ for various values of $F$ given in a.u. in the legend.}
    \label{fig:ioniz rate}
\end{figure}

Finally, the third term in \eqref{eq: lochfrass complex energy}, $\Gamma(R,F)$, is the strong-field ionization rate. This term directly accounts for the finite lifetime of the EGS in the presence of the field, mostly due to tunnel ionization. In a non-Hermitian approach it is simply implemented by including a non-zero imaginary part in the EGS energy. We show on \fref{fig:ioniz rate} the ionization rate $\Gamma(R,F)$ as a function of $R$ for a few field values \cite{labeye_tunnel_2018} (see section~\ref{app:numerical_details}).

We stress again that the internuclear distance $R$ and the time $t$ that appear in eqs.~\ref{eq: time-dependent dressed tise} and~\ref{eq: lochfrass complex energy} are only fixed parameters, and not variables. The fact that $R$ is a parameter is related to the BO approximation: the electrons instantaneously adapt to the nuclei. The fact that $t$ is a parameter is related to the adiabatic approximation: the electrons instantaneously adapt to the field. The total vibronic TDSE~\cite{lein_attosecond_2005,caillat2011a,caillat2018a}
\begin{align}
\label{eq: xR tdse}
i{\pdv{}{t}} \Psi (x,R,t)=\left[-\frac{1}{2\mu}{\pdv[2]{}{R}}+ V_{\mathrm{NN}}(R)+H_\mathrm{el}\right]\Psi(x,R,t)
\end{align}
can thus be reorganized as:
\begin{align}
\label{eq: BO factored tise}
i\widetilde{\varphi}_0(x;R,t){\pdv{}{t}} \chi(R,t)&=\widetilde{\varphi}_0(x;R,t)\Bigg[-\frac{1}{2\mu}{\pdv[2]{}{R}}+ V_{\mathrm{NN}}(R)+\varepsilon_0(R,F)\Bigg]\chi(R,t).
\end{align}
Note that we neglected the $\partial\widetilde{\varphi}_0(x;R,t)/\partial t$ term using the adiabatic approximation, and the $\partial^2\widetilde{\varphi}_0(x;R,t)/\partial R^2$ term using the BO approximation. We can divide~\eqref{eq: BO factored tise} by $\widetilde{\varphi}_0(x;R,t)$ to get a purely nuclear TDSE:
\begin{equation}
\label{eq:Lochfrass BO nuclear tdse}
\begin{split}
i{\pdv{}{t}}  \chi(R,t)=&\Bigg[-\frac{1}{2\mu}\pdv[2]{}{R} +V_{\mathrm{NN}}(R) + E_0(R) + \Delta E(R,F(t)) - i \frac{\Gamma(R,F(t))}{2}\Bigg]\chi(R,t).
\end{split}
\end{equation}
Therefore in this adiabatic BO formalism we completely separated the electronic and the nuclear dynamics. We can thus concentrate on the nuclear dynamics by considering a nuclear wave packet that evolves on a field-dressed PES. The interaction with the electron and with the laser field is taken into account solely through the instantaneous Stark shift~$\Delta E$ and strong-field ionization rate~$\Gamma$.

If one sets the Stark shift $\Delta E(R,F)$ to zero in \eqref{eq:Lochfrass BO nuclear tdse}, then we recover the equation of motion used in Ref.~\citenum{goll_formation_2006} to describe Lochfrass, here called the LF model. On the opposite, one might set the ionization rate $\Gamma$ to zero to describe only Bond-Softening, recovering what we called the BS model. However, setting one of these two terms to zero seems artificial, and thus, here we show the results of simulations including both contributions, which we called LBS.. We mention that this model is very close to the one used in Ref.~\citenum{varro_diatomic_2023}, where they give a fully analytical solution in the case of the Morse potential. Here we use a numerical approach, which we can use for any chosen potential.

\subsection{Fully correlated model}
\label{sec:num methods xR SP}
\begin{figure}[htp]
    \centering
    \includegraphics[width=0.48\linewidth]{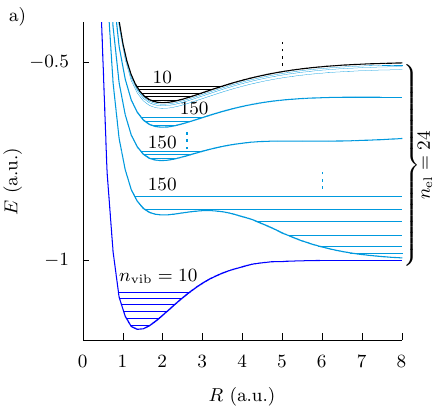}\hfill
    \includegraphics[width=0.48\linewidth]{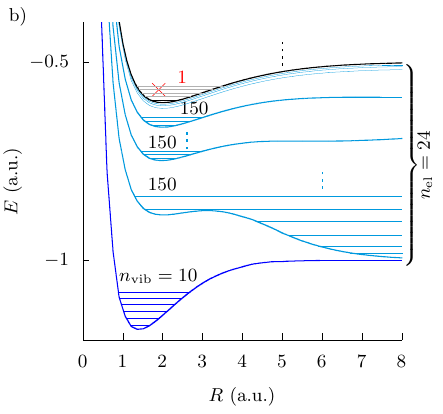}\\
    \includegraphics[width=0.48\linewidth]{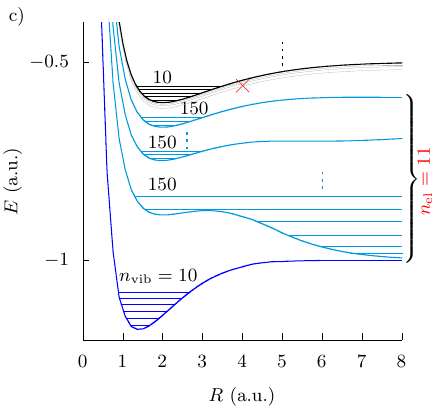}\hfill
    \includegraphics[width=0.48\linewidth]{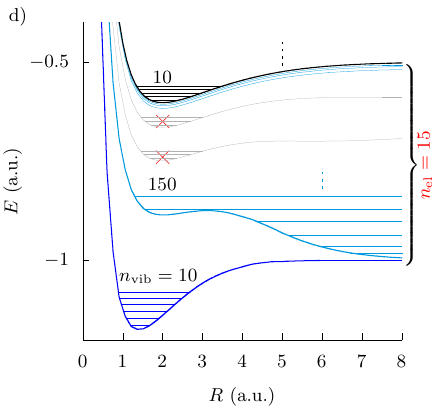}
    \caption{Schematic representation of the different active spaces used in the state-projected calculations\replaced{: $B_\mathrm{a}$ (a), $B_\mathrm{b}$ (b), $B_\mathrm{c}$ (c), $B_\mathrm{d}$ (d) (see text). The electronic ground state is in dark blue, the bound electronic states are in light blue, the lowest continuum electronic state is in black, vibrational states are depicted as horizontal lines. Basis $B_\mathrm{a}$ is the most complete one. The states included in $B_\mathrm{a}$, but not in the other bases are depicted in grey, with a red cross.}{ (see text).}}
    \label{fig:schema_FS}
\end{figure}

The fully correlated simulations treat both electronic and nuclear degrees of freedom at the same level of theory. We thus numerically solve the full vibronic TDSE (\eqref{eq: xR tdse}) to get the 2D time-dependent wave function.

This representation includes all the relevant neutral and ionic electronic states. To directly compare to the LBS model, and to what is measured in the experiments \cite{ergler_quantum-phase_2006,fang_strong-field_2008}, we compute the average value of $R$ within the EGS as
\begin{align}
\label{eq: average value R EGS}
\ev{R}_\mathrm{EGS}(t)=
\frac{\int R \left\lvert\chi_\mathrm{EGS}(R,t)\right\rvert^2 \dd{R}}{\int\left\lvert\chi_\mathrm{EGS}(R,t)\right\rvert^2\dd{R}},
\end{align}
where  
\begin{align}
\chi_\mathrm{EGS}(R,t)=\int\varphi_0(x;R)\Psi(x,R,t)\dd{x}
\end{align}
is the projection of the full time-dependent wave function $\Psi(x,R,t)$ onto the EGS $\varphi_0(x;R)$ of our system. Furthermore, we define the vibrational population of state $v$ as the projection of $\chi_\mathrm{EGS}$ onto the BO vibrational states $\chi_v$ of the the EGS $\varphi_0$, i.e.,
\begin{align}
\label{eq:population def}
    P_v(t)=\left|\int \chi_\mathrm{EGS}(R,t)\chi_v(R)\dd{R}\right|^2.
\end{align}
The phase of this projection can be used to extract the oscillation phase of $\ev{R}_\mathrm{EGS}(t)$ in the case where the population in the vibrational excited states of the EGS is dominated by the population in the first vibrational excited state:
\begin{align}
    \label{eq:phase from pop}
    \Phi_{10}=\phi_1(t_\mathrm{f})-\phi_0(t_\mathrm{f}) + \hbar\Delta E_\mathrm{vib} t_\mathrm{f}.
\end{align}
Here $t_\mathrm{f}$ is the time at the end of the pulse, and $\phi_v(t_\mathrm{f})$ is the phase of the projection $\int \chi_\mathrm{EGS}(R,t_\mathrm{f})\chi_v(R)\dd{R}$.

We use two different bases to represent the wave function: a two-dimensional grid and a ensemble of field-free Born-Oppenheimer states. The 2D grid results are easily checked for convergence with respect to all grid parameters, they are thus considered "exact" up to numerical accuracy and used as a reference throughout this work. We refer to these reference calculations as xR in the following. The BO basis is used to analyze the importance of the active space in the representation of the dynamics. We first found a sufficiently large basis to reproduce the exact results, and then we compared this "complete" active space to "restricted" active spaces, where some BO states were removed. The various BO bases used are sketched in \fref{fig:schema_FS}. We show the PES of several neutral and ionic electronic states. The total number of electronic states used in the neutral is called $n_\mathrm{el}$, and the total number of continuum electronic states in the ion is called $n_\mathrm{ion}$. For each electronic state $i$ we include a different number of vibrational states $n_\mathrm{nvib}(i)$. This number is written on top of the different PES in \fref{fig:schema_FS}. 

The largest basis, that we call $B_\mathrm{a}$, is depicted in \fref{fig:schema_FS}(a). It includes all neutral electronic states computed on the $x$ grid (see section~\ref{app:numerical_details}) corresponding to $n_\mathrm{el}=24$ neutral electronic states. It also includes all ionic continuum electronic states with an asymptotic kinetic energy $\epsilon$ below $\epsilon_\mathrm{max}\SI{=5.0}{a{.}u{.}}$, corresponding to $n_\mathrm{ion}=375$ ionic electronic states. We include $n_\mathrm{vib}(0)=10$ vibrational states for the neutral electronic ground state, $n_\mathrm{vib}(n\geq1)=150$ vibrational states for each neutral electronic excited state, and  for each ionic electronic state $n_\mathrm{vib}(\epsilon)=10$.
We compare this reference basis to three different ones. The basis $B_\mathrm{b}$, \fref{fig:schema_FS}(b), includes almost the same states as $B_\mathrm{a}$, except that only one vibrational state was included for each electronic continuum state. The basis $B_\mathrm{c}$, \fref{fig:schema_FS}(c), comprises almost the same states as $B_\mathrm{a}$, except that only the first 11 electronic states were included. The basis $B_\mathrm{d}$, \fref{fig:schema_FS}(d), is composed of almost the same states as $B_\mathrm{a}$, except that some electronic excited states are missing: the ground and first electronic excited states, and the 13 higher lying Rydberg states are included, but all electronic excited states in between are removed. Finally, we also checked the importance of the continuum active space by varying the maximum energy of the continuum states included. This is not depicted in \fref{fig:schema_FS}.

\subsection{Numerical details}
\label{app:numerical_details}
\subsubsection{Lochfrass and Bond Softening}
The Stark shift of our model systems (see eqs.~\ref{eq:Lochfrass molcoul},  \ref{eq:Lochfrass Coulomb} and \ref{eq: time-dependent dressed hamiltonian}) was computed with the so-called $R$-box method \cite{maier_spherical-box_1980} in one dimension. The strong-field ionization rate of our model systems was obtained by solving the \textit{electronic} one dimensional TDSE within a Born-Oppenheimer (BO) approach \cite{labeye_tunnel_2018}, over sampled values of the field $F$ and of the \textit{fixed} internuclear distance $R$, and then extrapolated for arbitrary $(R,F)$ values. No substantial difference in the dynamics was found when the analytical formula derived in \replaced{Ref.~\citenum{labeye_tunnel_2018}}{\cite{labeye_tunnel_2018}} was used for the strong-field ionization rate. Note that the 1D approach tends to overestimate the polarizability compared to the 3D case, which overestimates the Stark Shift, and underestimates the ionization rate. Since we are interested in qualitative behavior, rather than quantitative predictions, this does not affect our conclusions.

The nuclear wave packet is represented on a grid with 8000 points, separated by $\Delta R=2.5\times10^{-3}~\mathrm{a.u}$. The vibrational states are computed by numerical diagonalization of the field-free Hamiltonian with the LAPACK library. The ground vibrational state is then chosen as the initial state and propagated with a Crank-Nicolson \cite{crank_practical_1947} algorithm with 8192 times steps per laser cycle. 

In this representation, the expectation value $\ev{R}{\chi}$ directly yields the average value $\ev{R}_\mathrm{EGS}$ of the internuclear distance within the EGS of our model molecules. Finally, the population in the different vibrational states are computed by projection of the nuclear wave function on the eigenstates of the field-free nuclear Hamiltonian. 

\subsubsection{Grid basis}
The wave function is represented on a two-dimensional grid and the TDSE is solved with a split-operator algorithm:
\begin{align}
\ket{\Psi(t+\Delta t)} = &\exp[-iV\left(t+\frac{\Delta t}{2}\right)\frac{\Delta t}{2}]\exp[-iK\Delta t] \exp[-iV\left(t+\frac{\Delta t}{2}\right)\frac{\Delta t}{2}] \ket{\Psi(t)},
\end{align}
where $V(t)=V_\mathrm{NN}+V_\mathrm{eN}+H_\mathrm{int}(t)$ is the potential part of the Hamiltonian which only depends on the position operators $x$ and $R$, and is thus diagonal in the 2D grid basis, and  $K=\frac{P_R^2}{2\mu}+\frac{P_x^2}{2}$ is the kinetic energy operator. After multiplying $\ket{\Psi(t)}$ by the first exponential factor, we switch to momentum representation of the 2D wave function by 2D Fourier transform with the FFTW library \cite{frigo_design_2005}. Since $K$ is diagonal in this basis, we can easily apply the second exponential factor, then we go back to position representation by inverse Fourier transform and apply the third exponential factor to get $\ket{\Psi(t+\Delta t)}$.

The 2D grid has 4096 points in $x$ separated by $\Delta x=10^{-1}~\mathrm{a.u}$, and 1024 points in $R$ separated by $\Delta R=10^{-2}~\mathrm{a.u}$. We use 1024 time steps per laser cycle. The eigenstates of the correlated Hamiltonian, which corresponds to the full vibronic states, are computed by imaginary time propagation, and the ground state is taken as the initial state of the propagation. To avoid nonphysical reflections at the boundaries of the box, a $\SI{100}{a{.}u{.}}$ long $\cos^{1/8}$ mask type absorber \cite{krause_calculation_1992} is used in the $x$ direction. 

\subsubsection{Born-Oppenheimer basis}
The wave function is represented in the basis of field-free BO states. The BO electronic states $\varphi_i(x;R)$ and energies $\epsilon_i(R)$ are computed by diagonalizing the electronic Hamiltonian with the LAPACK library on a grid with 20001 points in $x$ separated by $\Delta x = 2\times10^{-2}~\mathrm{a.u}$. This diagonalization is performed for fixed values of $R$ ranging from 0 to $15~\mathrm{a.u}$, separated by a constant step of $0.15~\mathrm{a.u}$. The dipole matrix elements between electronic states are computed as
\begin{align}
d_{i,j}(R) = \int  \varphi_i(x;R)\,x\,\varphi_j(x;R)\,\mathrm{d}x
\end{align}
for all these values of $R$.

The electronic energies $\epsilon_i(R)$ and dipole matrix elements $d_{i,j}(R)$ are then linearly interpolated on a finer grid containing 8000 $R$ points ranging from 0 to $\SI{15}{a{.}u{.}}$ and then taken as constant. Within each electronic state $i$, the BO vibrational states $\chi^{(i)}_v(R)$ and energies $E^{(i)}_v$ are computed by diagonalizing the BO vibrational Hamiltonian on the finer $R$ grid. The BO dipole matrix elements $d^{(i,j)}_{v_1,v_2}$ are computed as $d^{(i,j)}_{v_1,v_2}=\int d_{i,j}(R)\chi^{(i)}_{v_1}(R)\chi^{(j)}_{v_2}(R)\mathrm{d}R$. These calculations were implemented using the Julia programming language \cite{bezanson_julia_2017}.

The field free Hamiltonian $H_0$ is approximated by a diagonal matrix containing the BO vibrational energies. The initial wave function is taken as the BO ground state, and propagated with a split-operator algorithm
\begin{align}
\ket{\Psi(t+\Delta t)} = &\exp[-iH_0\frac{\Delta t}{2}]\exp[-iF\left(t+\frac{\Delta t}{2}\right)X\Delta t]  \exp[-iH_0\frac{\Delta t}{2}] \ket{\Psi(t)},
\end{align}
where $F(t)$ is the electric field, $X$ is the dipole matrix and $\Delta t$ is the value of the time step. We used again 1024 time steps per laser cycle. 

\section{Results and discussion}
\label{sec:results}
Similarly to what was reported before \cite{ergler_quantum-phase_2006,goll_formation_2006}, we find that the strong infrared laser pulse triggers some nuclear dynamics in our model molecule, in both the parallel and the perpendicular cases. However, the different models that we detailed in the previous section are not all able to reproduce fully the quantitative and qualitative features that we observed. To analyze these differences in depth, we first compare the LBS model to the previously defined LF and BS models. Then we compare the LBS model to our fully correlated reference calculations, and finally we investigate the importance of the active space with the BO-basis calculations.

\subsection{Lochfrass and Bond-Softening}
\label{sec:results_BO}
\begin{figure}[htbp]
    \centering
    \includegraphics[width=0.7\linewidth]{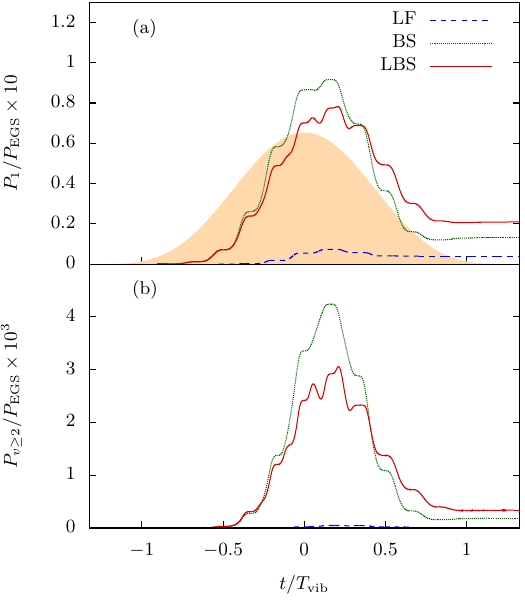}
    \caption{Population in the first vibrational states normalized by the population in the EGS of the neutral as a function of time for the parallel case. (a) Population in the first vibrational state of the EGS and (b) population in all higher lying vibrational excited states of the EGS. The field intensity envelope is displayed as an orange filled curve on panel (a).}
    \label{fig:poptimeLBS_H2}
\end{figure}

\begin{figure}[htbp]
    \centering
    \includegraphics[width=0.7\linewidth]{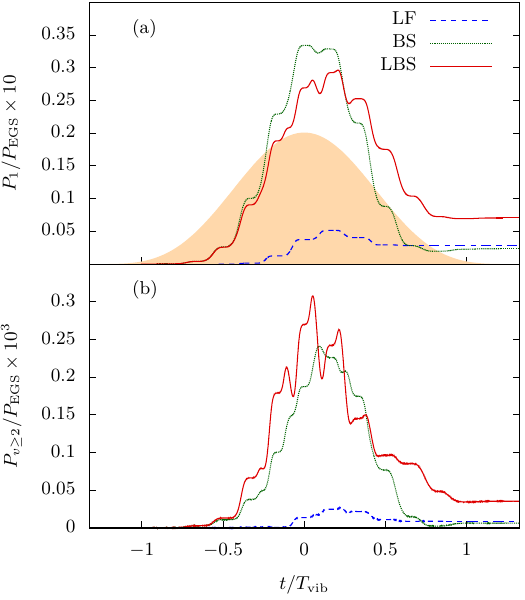}
    \caption{Population in the first vibrational states normalized by the population in the EGS of the neutral as a function of time for the perpendicular case. (a) Population in the first vibrational state of the EGS and (b) population in all higher lying vibrational excited states of the EGS. The field intensity envelope is displayed as an orange filled curve on panel (a).}
    \label{fig:poptimeLBS_A2}
\end{figure}

We start by comparing the differences and similarities between LF and BS. In both cases, the field triggers some nuclear dynamics in the EGS of the neutral model molecule. This indicates that some vibrational excited states of the neutral EGS are populated by the pulse. We show in \fref{fig:poptimeLBS_H2} and \fref{fig:poptimeLBS_A2} the population $P_v(t)$, see \eqref{eq:population def}, in the vibrational excited states of the neutral EGS as a function of time for the parallel and perpendicular case, respectively. We note that in the BS model, the time-dependant Hamiltonian is Hermitian, so that the norm of the wave function is conserved. However, when we include the ionization rate, the Hamiltonian is not Hermitian anymore: the norm of the wave function decreases with time to account for the population that leaves the EGS through strong-field ionization. To get a consistent comparison of these models, we thus renormalize the population in the vibrational states by the total population in the neutral EGS. In both \fref{fig:poptimeLBS_H2} and \ref{fig:poptimeLBS_A2}, the top panel (a) shows the population in the first vibrational excited state, and the bottom panel (b) shows the total population in the other vibrational excited states. 

We find that all mechanisms actually predict a quite similar bell-shaped behavior for the populations in the different vibrational states as a function of time. In agreement with Ref.~\citenum{ergler_quantum-phase_2006,goll_formation_2006}, we find that almost only the first vibrational excited state [\fref{fig:poptimeLBS_H2}(a) and \ref{fig:poptimeLBS_A2}(a)] gets populated. The population in the higher excited states [\fref{fig:poptimeLBS_H2}(b) and \ref{fig:poptimeLBS_A2}(b)] remains negligible and will have very little effect on the nuclear dynamics.

We see that the relative importance of the BS and LF mechanisms is different for the parallel and perpendicular cases. In the parallel case, BS induces a higher final population in the excited state than LF, and conversely for the perpendicular case. This is actually consistent with our previous observation that the Stark shift is strongly reduced in the perpendicular case (see section~\ref{sec:num_methods} and \fref{fig:stark_shift}), which reduces the effect of BS, and increases the effect of LF.

If we now compare the LF and BS models to the full LBS model, we see that the behavior is qualitatively the same: some population is transferred to the vibrational excited states, mainly in the first vibrational excited state. An important remark is that the final excited populations in the LBS model ($=0.021$) is not just a sum of the LF and BS models ($=0.017$). This indicates that LF and BS are not completely independent mechanisms, but that they are coupled. 

\begin{figure}[htbp]
    \centering
    \includegraphics[width=0.8\linewidth]{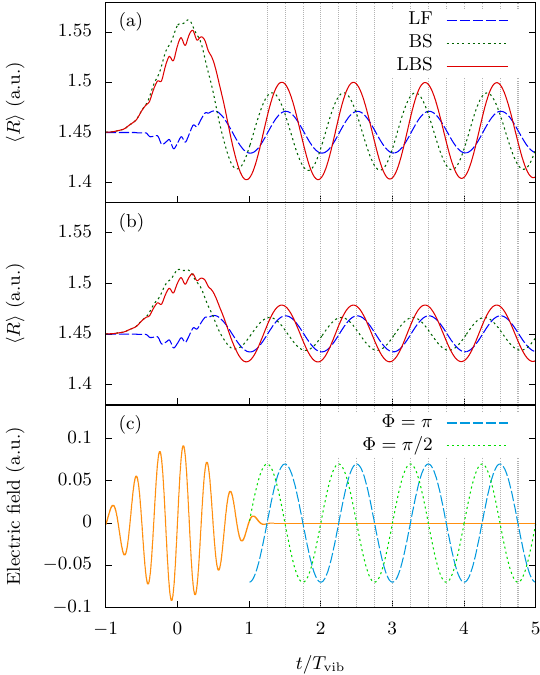}
    \caption{Average value of the internuclear distance in the EGS as a function of time for the (a) parallel and (b) perpendicular cases. Panel (c) shows the electric field as a function of time, and two sine functions with phase $\Phi=\pi$ and $\Phi=\pi/2$.}
    \label{fig:Rmean_BO}
\end{figure}

To investigate further, we show the average value of the internuclear distance $\ev{R}$ as a function of time, as calculated by these different models, in \fref{fig:Rmean_BO}. This is more directly related to what is actually measured in experiments \cite{ergler_quantum-phase_2006,fang_strong-field_2008}. As was found in these experimental works, we see that the internuclear distance oscillates, and, as soon as the field is turned off, behaves as
\begin{align}
    \ev{R}=R_0 + \delta_R\cos(\omega_\mathrm{vib} t - \Phi),
\end{align}
where $\omega_\mathrm{vib}$ is the energy difference between the ground and first vibrational excited state. In the parallel case (panel a), we find that the oscillation amplitude predicted by BS is larger than the one predicted by LF, while this is the contrary for the perpendicular case (panel b). This is perfectly consistent with the results shown in \fref{fig:poptimeLBS_H2} and \ref{fig:poptimeLBS_A2}.

The main difference between the predictions of the LF and BS models is the phase $\Phi$ of these oscillations. This was actually used experimentally to discriminate between the two mechanisms. In agreement with Ref.~\citenum{goll_formation_2006}, we find a phase close to $\pi$ for LF: $1.03\pi$, and $1.02\pi$ for the parallel and perpendicular case, respectively. However, the predictions of the BS model are not exactly equal to $\pi/2$ as claimed in Ref.~\citenum{goll_formation_2006}, we find $0.65\pi$ and $0.57\pi$ for the parallel and perpendicular case, respectively. 

The value of this phase was interpreted in Ref.~\citenum{goll_formation_2006} by the initial motion of the nuclei. In the case of Lochfrass, it depends on the slope of the ionization rate near the equilibrium internuclear distance. In the case of the \ce{H2} molecule, since the ionization potential decreases with $R$ close to the equilibrium distance, the strong-field ionization rate increases with $R$, so that Lochfrass initially drives the nuclear wave packet towards low values of the internuclear distance $R$. This is consistent with the results displayed in \fref{fig:Rmean_BO}: the average value $\ev{R}$ predicted by LF first decreases as a function of time, reaches a minimum at the maximum of the field envelope ($t=0$), and then oscillates. On the contrary, for Bond-Softening, since the polarization increases with increasing values of $R$, the BS mechanism loosen the molecular bond, and initially drives the nuclear wave packet towards larger values of $R$. This is also consistent with the results displayed in \fref{fig:Rmean_BO}: the average value $\ev{R}$ predicted by BS first increases as a function of time, reaches a maximum at the maximum of the field envelope ($t=0$), and then oscillates.

However, this interpretation only holds as long as we consider the LF and BS mechanisms separately. As we previously discussed, they can occur simultaneously and have to be included together. When we do so, as shown in \fref{fig:Rmean_BO}, we see that, in both the parallel and perpendicular case, the average value $\ev{R}$ first increases, as in the BS softening case, reaches a maximum at $t=0$, but then oscillates with a phase $\Phi=0.92\pi$ which is much closer to the LF predicted one. This contradicts the interpretation that the oscillation phase is controlled by the initial motion of the nuclear wave packet. As a consequence, our results indicate that the slope of the ionization rate near the equilibrium distance cannot directly be deduced from this oscillation phase~$\Phi$, as was claimed in Ref.~\citenum{fang_strong-field_2008}. We emphasize that our results do not infer the claim that, in the cases where LF is dominating, the slope of the ionization rate controls the initial motion of the nuclear wave packet. However, we can conclude that this initial motion itself does not always determine the oscillation phase.

Moreover, in both the parallel and the perpendicular case, we find a phase $\Phi=0.92\pi$, which is close to the Lochfrass prediction, even though the BS mechanism is predominant in the parallel case. This strongly challenges the interpretation that the oscillation phase allows to discriminate between the two mechanisms. On the contrary, our results indicate that both mechanisms are actually coupled, and cannot be seen as independent from one another. The oscillation phase is a consequence of the coupled dynamics, and not of the relative importance of one mechanism with respect to the other. In all the following, we will thus only consider the full LBS model, and compare it to reference fully correlated models. To simplify the discussion, we focus on the aligned case.

\subsection{Correlations}
\begin{figure}[htbp]
\centering
\includegraphics[width=0.8\linewidth]{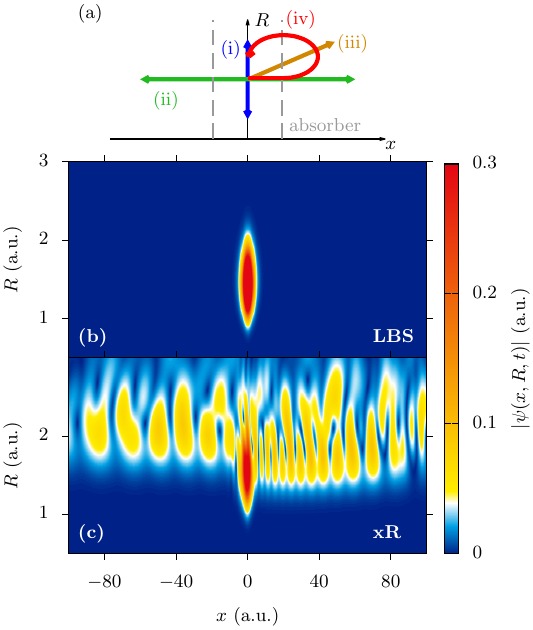}
\caption{Illustration of the vibronic correlations in the $(x,R)$~plane. (a) Schematic representation of the different contributions to the vibronic dynamics (i) uncorrelated nuclear dynamics, (ii) uncorrelated electronic dynamics, (iii) correlated dynamics, (iv) strongly correlated dynamics. Absolute value of the wave function at $t=0.27\times T_{\mathrm{vib}}$ for a laser pulse with~$\CEP=\pi/2$ computed with (b) the LBS model and (c) the fully correlated simulations.}
\label{fig:lochfrass correlations}
\end{figure}

We now want to investigate more in depth the capabilities and limits of the LBS model. For this, we represented in \fref{fig:lochfrass correlations}(a) the different levels of vibronic correlation to be considered.  In particular, since LBS relies on the BO approximation, the vibronic correlations between the electron and the nuclei cannot be perfectly represented in this model. These vibronic correlations can originate from different contributions. The dynamics that are confined along the~$R$~axis, (i), or along the~$x$~axis, (ii), are \textit{uncorrelated}. We expect them to be well represented by a factorized time-dependent wave function as in \eqref{eq: BO time-dependent wave function}. Conversely, the diagonal contributions, sketched as (iii) and (iv), couple the electronic and nuclear degrees of freedom, either when an electron leaves the nuclei but never comes back (iii), or when an electron is excited or ionized but interferes later on with the nuclei (iv). These correlations should not, a priori, be well represented by a factorized wave function. Nevertheless, the strength of the LBS formalism is that their \textit{imprint on the nuclear dynamics} may itself be included. This is actually handled by the Stark shift~$\Delta E$ and strong-field ionization rate~$\Gamma$ in \eqref{eq:Lochfrass BO nuclear tdse}. Indeed, these two electronic terms depend on the nuclear coordinate $R$, and will thus affect the nuclear dynamics. The  ionization rate~$\Gamma$ allows to represent dynamics where an electron is ionized and never comes back to the nuclei afterwards. It therefore allows to represent the correlations sketched as (iii) in \fref{fig:lochfrass correlations}. The effect of the Stark shift term~$\Delta E$ is somewhat less clear in this $(x,R)$ picture, but it will mostly couple the electronic states of the neutral molecule, so that it will allow to represent dynamics where the electron stays close to the nuclei.

To summarize, we expect the LBS model to be able to account for dynamics that are either decoupled, (i) (ii); coupled dynamics where the electron stays in a bound electronic state; or coupled dynamics where the electron is ionized, leaves the nuclei and do not come back, (iii). Consequently, if we had some dynamics where, in the idea of the renown three-step model \cite{corkum_plasma_1993,schafer_above_1993,lewenstein_theory_1994}, an electron would be ionized, then, upon acceleration by the laser field, would be brought back to the nuclei, and there would recombine to the ground state, these "trajectories" would be completely absent from the LBS model. Even their \textit{imprint} on the nuclear dynamics could not be represented properly. There contributions are sketched as (iv) in \fref{fig:lochfrass correlations}. We assess their impact on the nuclear dynamics in the following by confronting the LBS results with the fully correlated simulations.

We stress that it is very difficult to disentangle the contributions that we just mentioned directly from the complete time-dependent wave function. This is illustrated on panels (b) and (c) of \fref{fig:lochfrass correlations}, where we show the wave function at a representative time $t\simeq0.27\times T_\mathrm{vib}$ during the laser pulse computed either in the LBS formalism (b), or with the fully correlated xR model (c). As expected we see that the LBS wave function can only represent the nuclear dynamics close to the $R$~axis, while the xR model shows complex interferencing wave function. In order to discriminate the various levels of correlation in the fully vibronic wave packet $\psi(x,R,t)$, we use a numerical absorber, as illustrated in \fref{fig:lochfrass correlations}(a), to discard the strongly correlated contributions (iv) before the electron returns to interfere with the EGS. From the classical three-step model, we expect the maximum excursion to be about $x_\mathrm{max}=2F_0/\omega_\mathrm{L}^2\SI{=57}{a{.}u{.}}$ for the "long trajectories", and half as much for the "short trajectories" \cite{schafer_above_1993,corkum_plasma_1993}. We therefore place our numerical absorber closer than this value to get an effective absorption of these trajectories.

\begin{figure}[htbp]
\centering
\includegraphics[width=0.8\linewidth]{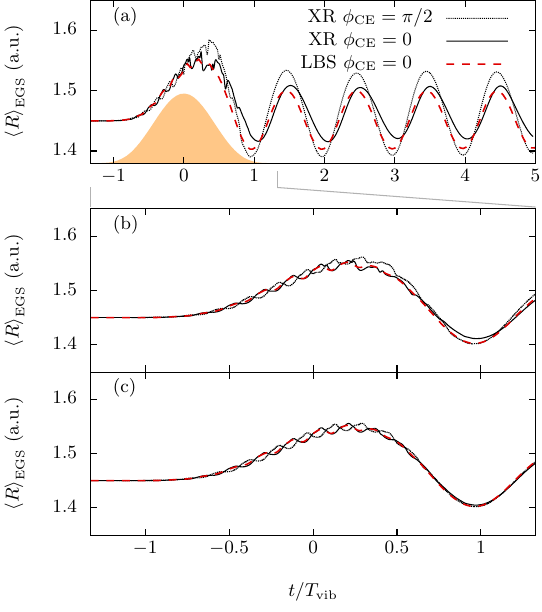}
\caption{Average value of the internuclear distance in the electronic ground state as a function of time for the parallel case. Results obtained with the LBS and xR models for different values of the CEP (see legend). The xR simulations are performed with an absorber acting from (a) $x_\mathrm{abs}\SI{=104.8}{a{.}u{.}}$, (b) $x_\mathrm{abs}\SI{=28.46}{a{.}u{.}}$, (c) $x_\mathrm{abs}\SI{=15}{a{.}u{.}}$ from the center of mass of the molecule. The time frame in panels (b) and (c) is restricted to the support of the pulse envelope, displayed as an orange filled curve in panel (a).}
\label{fig:Lochfrass abs xRBO}
\end{figure}

In \fref{fig:Lochfrass abs xRBO}, we plot the average value $\ev{R}_\mathrm{EGS}(t)$, see \eqref{eq: average value R EGS}, as it evolves in time, with different absorbing conditions. In panel (a) the absorber is located at $x_\mathrm{abs}=\SI{104.8}{a{.}u{.}}\simeq1.8\times x_\mathrm{max}$, i.e., sufficiently far away from the molecule so that it does not affect the bound state dynamics. 
We observe here that both the LBS and xR simulations predict a similar oscillatory behavior with a near $\pi$ phase. However, when we compare the results of the LBS with the xR model for $\CEP=0$, we observe several discrepancies, indicating that vibronic correlations indeed affect the nuclear dynamics. We see that the two models noticeably predict different values of the phase, amplitude, and average value of the oscillations. This indicates that the predicted nuclear wave packets are different both in terms of vibrational state populations and phases. But the most remarkable difference is that the results predicted by the xR model unexpectedly depend on the carrier envelope phase of the laser pulse \CEP. This feature is completely absent from the LBS predictions, (not shown in \fref{fig:Lochfrass abs xRBO}, but in the following figures).

To understand which of the correlations discussed above is responsible for these discrepancies, we move our numerical absorber closer to the molecule. Panels (b) and (c) of \fref{fig:Lochfrass abs xRBO} display the results obtained with $x_\mathrm{abs}=\SI{28.5}{a{.}u{.}}=0.5\times x_\mathrm{max}$ and $x_\mathrm{abs}\SI{=15}{a{.}u{.}}=0.26\times x_\mathrm{max}$, respectively. We clearly see that the agreement between the two models increases when the absorber gets closer. On panel (c), where the absorber is so close that almost all ionized parts of the wave function are removed, we even reach a perfect agreement between the LBS (uncorrelated) and xR (correlated) models. This strongly suggests that the discrepancies between the two models are actually due to dynamics where the electron leaves the EGS and undergo recollision later on [path (iv) sketched in \fref{fig:lochfrass correlations}(a)]. The LBS is by essence unable to describe such a mechanism since it only treats the dynamics in the EGS, and discards the ionized part of the wave function. This explains why we get a perfect agreement with the LBS model when we absorb this part of the wave function before it can come back close the nuclei, thus preventing any kind of rescattering. The numerical absorber in position representation used in the xR model thus plays a similar role as the "absorbing" imaginary potential term $-i\Gamma/2$ in the nuclear TDSE (\eqref{eq: BO factored tise}). 

\begin{figure}[htbp]
\centering
\includegraphics[width=0.8\linewidth]{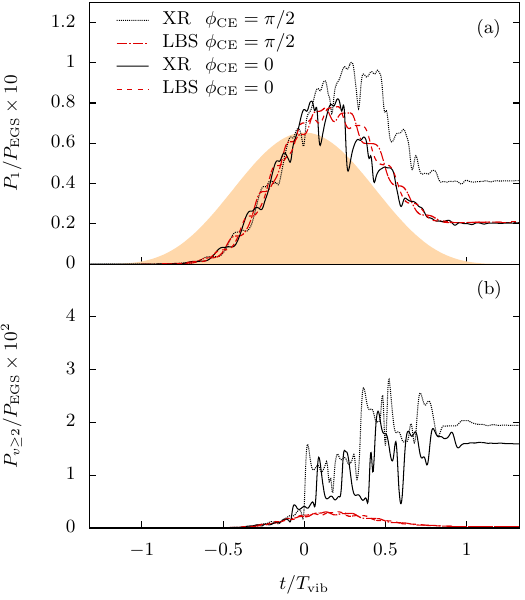}
\caption{Population in the first vibrational states normalized by the population in the EGS of the neutral as a function of time for the parallel case. (a) Population in the first vibrational state of the EGS and (b) population in all higher lying vibrational excited states of the EGS. Results obtained with the LBS and xR models for different values of the CEP $\CEP=0$ and $\CEP=\pi/2$. The field intensity envelope is displayed as an orange filled curve on panel (a).}
\label{fig:poptime XR LBS}
\end{figure}
 
\subsection{CEP dependence}
An interesting feature in \fref{fig:Lochfrass abs xRBO}(a) is the difference in the nuclear dynamics predicted by the fully correlated xR model for two different values of the CEP, $\CEP=0$ and $\pi/2$. This feature, that has, to our knowledge, not been identified nor commented before, is completely absent from the LBS model. We investigate this behavior in this section.

We plot the normalized population in the vibrational excited states of the EGS as a function of time in \fref{fig:poptime XR LBS}. We see that population in the first vibrational excited state (a) has a step-like evolution that adiabatically follows the successive maxima of the field on the rising front of the pulse. The curves corresponding to a CEP of 0 and $\pi/2$ are thus, as expected, out of phase, but have the same average envelope for times $t<0$. With the LBS model, the two curves continue to follow each other until the end of the laser pulse. However, in the xR simulations, we see that the two curves depart from each other after the pulse maximum ($t=0$). They reach significantly different final populations after the pulse end ($t\geq 1.32 \times T_\mathrm{vib}$). This leads to different $\ev{R}_\mathrm{EGS}$ oscillation amplitudes -- which are directly proportional to the population in the first vibrational excited state $P_1$ (eq.\ref{eq:population def})-- as observed in \fref{fig:Lochfrass abs xRBO}(a). The total population in the higher lying vibrational states is shown in \fref{fig:poptime XR LBS}(b), we see that the fully correlated xR simulations predict a much larger population in the higher lying vibrational excited states. This emphasizes the importance of the vibronic correlations influence on the nuclear dynamics. Note, however, that these results are not to be over-interpreted: the population in the different states is actually gauge-dependent as long as the field is not zero. To avoid this problem, in the following, we concentrate on the population at the end of the pulse, when the field is zero.

\begin{figure}[htbp]
\centering
\includegraphics[width=0.8\linewidth]{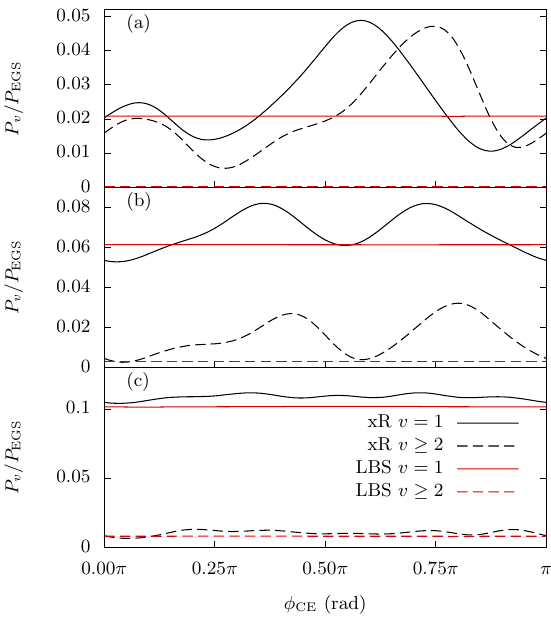}
\caption{Population in the first vibrational states $v$ of the molecule at the end of the pulse, normalized by the overall population in the EGS, as a function of \CEP for a laser duration of (a) 8 optical cycles, (b) 6 optical cycles and (c) 4 optical cycles. Results obtained with the LBS and xR models. }
\label{fig:Lochfrass CEP xRBO}
\end{figure}

\begin{figure}[htbp]
\centering
\includegraphics[width=0.8\linewidth]{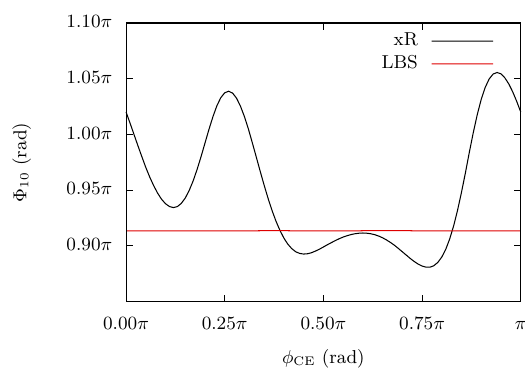}
\caption{Oscillation phase of $\ev{R}_\mathrm{EGS}(t)$ as a function of CEP, extracted by projecting the final wave function on the field-free first and ground vibrational states of the EGS (see \eqref{eq:phase from pop}). Results obtained with the LBS and xR models.}
\label{fig:Lochfrass phase CEP xRBO}
\end{figure}

To characterize this effect more precisely, we show in \fref{fig:Lochfrass CEP xRBO} the population in the first few vibrational excited states $v$ of the EGS at the end of laser pulse as a function of the CEP \CEP for different pulses duration. Panel (a) corresponds to the same laser conditions as \fref{fig:Lochfrass abs xRBO}. These results clearly show that the LBS nuclear dynamics does not have any CEP dependence. On the contrary, the xR populations display strong variations with the CEP. We confirm this trend when looking at the oscillation phase $\Phi$ of $\ev{R}_\mathrm{EGS}(t)$, which is plotted as a function of \CEP on \fref{fig:Lochfrass phase CEP xRBO}. We see on this figure that the oscillation phase predicted by the xR simulations also depends on \CEP, while the phase predicted by LBS is constant. We thus believe that the CEP is a promising parameter knob to selectively observe the effects of the vibronic correlations on the nuclear dynamics.

We then tried to identify a proper way to observe signatures of this \CEP dependence on the nuclear dynamics. Intuitively, one would think that a CEP-dependence is directly related to the laser pulse duration. Indeed, one expects the CEP dependence to be averaged out for longer pulses \cite{chelkowski_phase-dependent_2004,gurtler_asymmetry_2004,roudnev_general_2007,dong_unraveling_2024}. However it unexpectedly turns out  that the \CEP dependence gets even more pronounced when the pulse duration increases, as clearly seen in \fref{fig:Lochfrass CEP xRBO}. The effects of the vibronic correlations rather seem to build up during the laser pulse, requiring longer pulses to have a substantial influence.

\subsection{Importance of the active space}
Eventually, we further investigated the mechanisms underlying these correlated processes by progressively restricting the active space in the simulation, in the spectral domain rather than in the physical $(x,R)$ plane as previously. As detailed previously, we performed simulations where we solved the full TDSE (\eqref{eq: xR tdse}) by expanding the wave function $\psi(x,R,t)$ over several sets of BO states, spanning different energy regions of the Hilbert space. We consider a reference BO basis $B_a$, described in detail in the Numerical methods and sketched in \fref{fig:schema_FS}(a) whose results match the ones obtained with the grid xR simulations, and similar bases but where some degree of freedom is restricted.

\begin{figure}[htbp]
\centering
\includegraphics[width=0.7\linewidth]{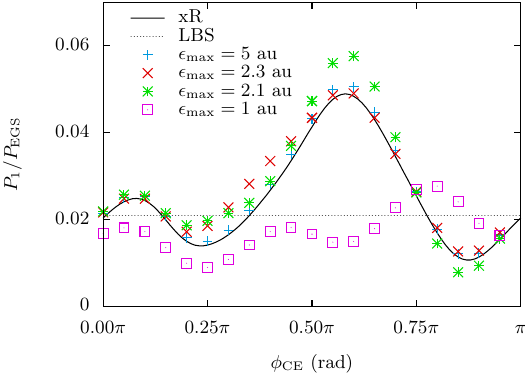}
\caption{Normalized population in the first vibrational excited state of the EGS at the end of the pulse as a function of CEP. The BO basis used is the full basis depicted in \fref{fig:schema_FS}(a), but with different values of $\epsilon_\mathrm{max}$ (see legend).}
\label{fig:SP cont}
\end{figure}

We first investigate the importance of the continuum representation by comparing the basis $B_a$ to BO bases almost equivalent, but with different number of electronic continuum states. The basis $B_a$ contains all continuum states up to an electron kinetic energy of $\epsilon_\mathrm{max}=\SI{5}{a{.}u{.}}=\SI{136}{eV}=7.6U_\mathrm{p}$ where $U_\mathrm{p}=F_0^2/4\omega_\mathrm{L}^2$ is the peak ponderomotive potential of the pulse. This corresponds to 375 continuum states. The other bases have $\epsilon_\mathrm{max}=\SI{2.3}{a{.}u{.}}=\SI{62.6}{eV}=3.5U_\mathrm{p}$, $\epsilon_\mathrm{max}=\SI{2.1}{a{.}u{.}}=\SI{47.1}{eV}=3.2U_\mathrm{p}$ and $\epsilon_\mathrm{max}=\SI{1.0}{a{.}u{.}}=\SI{27.2}{eV}=1.5U_\mathrm{p}$ with 247, 235 and 156 continuum states, respectively. We show in \fref{fig:SP cont} the final normalized population in the first vibrational excited state as a function of CEP for $B_a$ and these three other bases. We see that the results obtained with the basis $B_a$ perfectly match the exact xR results. As expected, the results obtained with the other bases gets closer to the exact ones when $\epsilon_\mathrm{max}$ is increased, i.e., when the basis size is increased. More precisely, we observe the results get much better as soon as $\epsilon_\mathrm{max}$ gets higher than the classical limit $\epsilon_\mathrm{max}\simeq 3.2\times U_\mathrm{p}$, which corresponds to the maximum kinetic energy of electrons that gets rescattered towards the nuclei \cite{corkum_plasma_1993,schafer_above_1993}, and is reminiscent of the cutoff energy of High-order Harmonic Generation (HHG) spectra \cite{corkum_plasma_1993,schafer_above_1993,lewenstein_theory_1994}. 

This confirms our interpretation that the nuclear dynamics is affected by parts of the wave packet that come back to the nuclei after ionization. Our results are consistent with Ref.~\citenum{monfared_influence_2022} where the nuclear dynamics was shown to impact the HHG spectra of diatomic molecules, but in a complementary way, we show that the HHG-like trajectories have an impact on the nuclear dynamics.

\begin{figure}[htbp]
\centering
\includegraphics[width=0.7\linewidth]{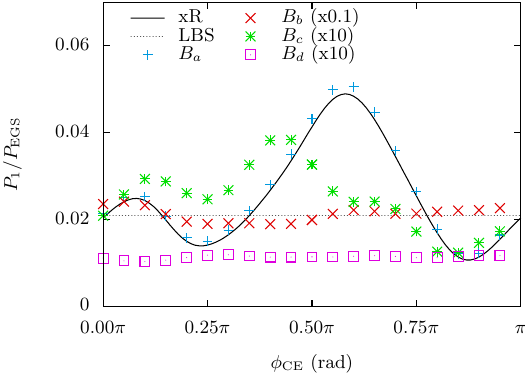}
\caption{Normalized population in the first vibrational excited state of the EGS at the end of the pulse as a function of CEP. The different BO basis used are described in the Numerical methods and sketched in \fref{fig:schema_FS}.}
\label{fig:SP bound}
\end{figure}

We also checked that not only the ionic electronic states are important, but the ionic nuclear vibrational states: we compare the reference basis $B_a$ with a basis $B_b$ where we removed all ionic vibrational excited states [see \fref{fig:schema_FS}(b)]. The results are shown in \fref{fig:SP bound}. We see that this restricted basis $B_b$ is not able to reproduce the correct nuclear dynamics. The results are independent of the CEP, indicating that the vibronic correlations seem to be absent in this case. Moreover, the population in the first vibrational excited state is largely overestimated.

After the continuum states, we also investigate the importance of the electronic bound states to reproduce the nuclear dynamics. We compare the reference basis $B_a$ to a basis $B_c$ (no Rydberg states), and a basis $B_d$ keeping only the ground and first electronic excited state and the Rydberg states (see \fref{fig:schema_FS}). The results are shown in \fref{fig:SP bound}. We see that neither bases $B_c$ nor $B_d$ are able to reproduce the dynamics. In both cases, almost no population gets transferred to the first vibrational excited state. This indicates that the electronic dynamics needs to be accurately represented to correctly reproduce the nuclear dynamics. 

All these results enforce our interpretation that the CEP dependence in the dynamics indeed stems from strong correlations between electronic and nuclear degrees of freedom. Both bound and continuum electronic states have to be well represented to properly model the considered strong-field driven molecular dynamics, somehow reminiscent of a \textit{chirp-dependent} adiabatic passage~\cite{saamlann2018a}, which cannot be fully accounted for by the models currently available.

\section{Conclusion}
To summarize, we investigated the strong field driven ultrafast nuclear dynamics of a model diatomic molecule. We confronted different adiabatic models based on the Born-Oppenheimer approximation: Lochfrass, Bond-Softening and a combination of both. We showed that since Lochfrass and Bond-Softening happen simultaneously, they can be coupled and should therefore be treated in a common formalism. We also challenged the previous interpretation that the phase of the molecular bond length oscillations was directly related to the initial motion of the nuclei, and eventually could be used to discriminate between Lochfrass and Bond-Softening. Conversely, our results indicate that the phase can sometimes be independent of the relative importance of these two mechanisms. More explicitly, if either LF or BS is clearly dominating, then this mechanism will control the oscillation phase (respectively $\pi$ or $\pi/2$ in the conditions of this work). However the reciprocal is not true: a given value of the phase does not allow to conclude on the dominating mechanism. In particular, a $\pi$ phase does not imply that Lochfrass is clearly dominating. We expect these conclusions to hold for other peak intensities and pulse duration, and also in the case of no CEP control.

We also analyze in details how vibronic correlations can affect the nuclear dynamics. We were able to discriminate between two different kinds of correlations. The first one is related to electronic excitation and direct ionization. It can be perfectly reproduced even when nuclear and electronic dynamics are considered separately, and is therefore well accounted for by the Lochfrass and Bond-Softening mechanisms. The second one originates from recollision processes and has, to our knowledge, never been described before. Our results established that the latter is responsible for a CEP dependence of the field induced nuclear dynamics in the EGS, that is absent from the Lochfrass and Bond-Softening mechanisms. We emphasize that, even though our results were obtained with low-dimensional model systems, and are therefore not quantitatively comparable to an experiment, the qualitative description of the processes described should remain valid. Further analytical developments and CEP-resolved experiments are needed to fully interpret the CEP dependence of the nuclear dynamics evidenced in the present work, with the perspective of gaining a deeper insight on vibronic correlations driven by strong fields.

\begin{acknowledgements}
We acknowledge financial support from the LABEX Plas@Par-ANR-11-IDEX-0004-02,  ANR-15-CE30-0001-01-CIMBAAD and ANR-20-CE30- 655 0007-DECAP.
\end{acknowledgements}

\bibliography{biblio}

\end{document}